\documentclass[usenatbib,usegraphicx,useAMS]{mn2e}
\usepackage{amsmath}
\usepackage{amssymb}

\def\cf{{cf.}} 
\def\eg{{e.g.}} 
\def\ie{{i.e.}} 

\def\s{{\rm s}} 

\def\GHz{{\rm GHz}} 

\def\m{{\rm m}} 
\def\cm{{\rm c}\m} 

\def\G{{\rm G}} 

\def\muas{\mu{\rm as}} 

\newcommand{\SgrA}{Sgr A*}
\newcommand{\Msun}{{{\rm M}_\odot}}

\def\figfactor{0.9}

\renewcommand{\d}{{\rm d}}

\title[Imaging Spots in Black Hole Accretion Flows]{Imaging
  Optically-Thin Hot Spots Near the Black Hole Horizon of \SgrA~at
  Radio and Near Infrared Wavelengths}
\author[Avery E. Broderick \& Abraham Loeb]{Avery
E. Broderick\thanks{E-mail: abroderick@cfa.harvard.edu} \& Abraham
Loeb\thanks{E-mail: aloeb@cfa.harvard.edu}\\ Institute for Theory and
Computation, Harvard-Smithsonian Center for Astrophysics, 60 Garden St., MS
51, Cambridge, MA 02138, USA\\}

\begin{document}
\maketitle

\begin{abstract}
Sub milli-arcsecond astrometry and imaging of the black hole \SgrA~at the
Galactic centre may become possible in the near future at infrared and
sub-millimetre wavelengths.  Motivated by observations of short-term
infrared and X-ray variability of \SgrA, in a previous paper we computed the
expected images and light curves, including polarization, associated with
an compact emission region orbiting the central black hole.
We extend this work, using a more realistic hot-spot model and including
the effects of opacity in the underlying accretion flow.  We find that at
infrared wavelengths the qualitative features identified by our
earlier work are present, namely it is possible to extract the
black hole mass and spin from spot images and light curves of the observed
flux and polarization.  At radio wavelengths, disk opacity produces
significant departures from the infrared behaviour, but there are still
generic signatures of the black hole properties.  Detailed comparison of
these results with future data can be used to test general relativity and
to improve existing models for the accretion flow in the immediate vicinity
of the black hole.
\end{abstract}


\section{Introduction} \label{I}

Testing general relativity in a regime where spacetime curvature is large
remains one of the primary goals of observational astronomy.  Black holes
provide a natural target in which to focus these efforts.  Nonetheless, due
to their extremely compact nature, an unambiguous signature of strong
gravity has been elusive so far.

A number of researchers have suggested that imaging an optically-thin
background accretion flow could provide a direct test of strong gravity
\citep[see, \eg,][]{Falc-Meli-Agol:00,Taka:05,Taka:04,Brod-Loeb:05b}.  The
angular scale of the black hole in the Galactic centre (identified with the
radio source \SgrA) is $5$-$10\,\muas$, twice as large as the nuclear black
hole in M87, and substantially larger than all other candidates.  As a
result, \SgrA~is the most promising candidate for high resolution imaging.
Within the next decade it is expected that a Very Long Baseline Array
(VLBA) of
sub-millimetre telescopes will exist, providing $20\,\muas$ resolution
imaging capabilities \citep{Doel-Bowe:04,Miyo_etal:04}.  Thus, theoretical
efforts to compute the images of realistic accretion flows is warranted.
In \citet{Brod-Loeb:05b} it was shown that for a typical accretion model,
opacity will be significant at radio wavelengths, substantially altering
the shape and visibility of the black hole ``shadow'' cast by the black
hole's photon capture cross-section.

Multi-wavelength polarization observations of accreting black holes have
also been proposed as a method of probing the black hole vicinity
\citep{Brod-Loeb:05b,Conn-Star:80,Laor-Netz-Pira:90}.
For optically-thin but geometrically-thick accretion flows, as expected in
low luminosity galactic nuclei such as our Galactic centre, it was shown by
\citet{Brod-Loeb:05b} that comparisons of the polarization near and above
the optically thick/thin transition frequency is indicative of the
black hole spin.

However, observations of near-infrared (NIR) and X-ray flaring of
\SgrA~have implied that the inner regions of the accretion flow are
nonuniform
\citep{Baga_etal:01,Genz_etal:03,Gold_etal:03,Ecka_etal:04,Ghez_etal:04}.
The time scale of the variability, $\sim10\,\min$, is comparable to the
period of the innermost stable circular orbit (ISCO), and thus suggestive
of an orbiting hot spot.  Therefore, it is likely that modeling of
\SgrA~images will require the inclusion of variability.  In
\citet{Brod-Loeb:05} it was shown that the images and light curves
associated with a hot spot could be used to measure the mass and spin of
the black hole.
There are additional plans to use the Phase Referenced Imaging and
Astrometry (PRIMA) instrument, currently under
construction, at the Very Large Telescope (VLT) to obtain
sub-milli-arcsecond astrometry of these flares \citep{Paum_etal:05}.

Here we improve upon the calculations of \citet{Brod-Loeb:05} by making use
of a more realistic hot spot model consisting of a localized population of
non-thermal electrons (which are produced, for example, by a magnetic
reconnection flare).  In addition, we include the effects of disk opacity
for a typical radiatively inefficient accretion flow model.
The opacity systematically decreases the
magnification and polarization fraction, smoothes out the polarization
angle light curve, circularizes the motion of the centroid, and symmetrizes
the time averaged spot images.

Section \ref{CM} presents a summary of the computational methods used and
the models for the accretion disk and hot spot.  Sections \ref{NIR} and
\ref{R} contain the light curves and centroid paths for NIR and
sub-millimetre frequencies, respectively.  The effects of opacity are
highlighted in section \ref{GEoO}.  Finally, section \ref{C} summarizes our
conclusions.

In what follows, the metric signature is taken to be \mbox{$(-+++)$},
and geometrised unites are used ($G=c=1$).

\section{Computational Methods} \label{CM}

\subsection{Ray Tracing \& Radiative Transfer} \label{CM:RTRT}
The method by which the light rays were produced and the radiative
transfer performed is discussed at length in \citet{Brod-Blan:03} (for
which tracing null geodesics is a limiting case) and
\citet{Brod-Blan:04} \citep[see also][]{Brod:05}, respectively.  As
such, only a brief summary is presented here.

Null geodesics are constructed by integrating the equations
\begin{equation}
\frac{\d x^\mu}{\d\lambda} = f(r) k^\mu
\qquad
{\rm and}
\qquad
\frac{\d k_\mu}{\d\lambda} = - f(r)
\left(\frac{1}{2}\frac{\partial k^\nu k_\nu}{\partial x^\mu}\right)_{k_\alpha}\,,
\end{equation}
where the partial differentiation is taken holding the covariant
components of the wave four-vector, $k_\mu$, constant.  The function
$f(r)$ is arbitrary (corresponding to the freedom inherent in the
affine parameterization, $\lambda$) and chosen to be
\begin{equation}
f(r) = r^2 \sqrt{1 - \frac{r_h}{r}}\,,
\end{equation}
(where $r_h$ is the horizon radius) in order to regularize the affine
parameter near the horizon.  \citet{Brod-Blan:03} have explicitly
demonstrated that this procedure does produce the null geodesics.
 
Polarized radiative transfer in curved spacetime is most easily
performed by integrating the Boltzmann equation
\citep{Lind:66,Brod:05}.  In this case, it is the photon distribution
function $N_\nu\propto I_\nu/\nu^3$ that is evolved.  In the case of
polarized radiative transfer, it is possible to define covariant
analogues of the Stokes parameters, $\bmath{N}_\nu=(N_\nu,N^Q_\nu,N^U_\nu,N^V_\nu)$
\citep{Brod-Blan:04}.  In terms of these, the radiative transfer
equation is
\begin{equation}
\frac{\d \bmath{N}_\nu }{\d\lambda}
=
\bar{\bmath{j}}_\nu
-
\bar{\bmath{\alpha}}_\nu \bmath{N}_\nu\,,
\end{equation}
where $\bar{\bmath{j}}_\nu$ and $\bar{\bmath{\alpha}}_\nu$ are emissivity
and absorption coefficients \citep[to see how these are related to their
standard definitions, see][]{Brod-Blan:04,Lind:66}.  The two synchrotron
emission components considered here involve populations of electrons with a
thermal \citep{Yuan-Quat-Nara:03} and power-law \citep{Jone-ODel:77a}
distributions.  Because we are primarily concerned with emission at
sub-millimetre or shorter wavelengths, Faraday rotation and conversion are
unlikely to be important, and are thus neglected.

\subsection{Disk Models} \label{CM:DM}
\begin{table}
\begin{center}
\begin{tabular}{ccccc}
\hline
{$a\,(M)$} & {$n_e^0\,(\cm^{-3})$} & {$T_e^0\,(K)$}
& {$n_{\rm nth}^0\,(\cm^{-3})$} & {$p_{\rm nth}$}\\
\hline
$0$ & $3\times10^7$ & $1.7\times10^{11}$ &
$8\times10^4$ & $-2.9$\\
$0.5$ & $3\times10^7$ & $1.4\times10^{11}$ &
$5\times10^4$ & $-2.8$\\
$0.998$ & $1\times10^7$ & $1.5\times10^{11}$ &
$1\times10^5$ & $-2.8$\\
\hline
\end{tabular}
\end{center}
\caption{The accretion flow parameters associated with equations
  \ref{model_eqs}.}
\label{model_params}
\end{table}
We use the same three accretion flow models as described in
\citet{Brod-Loeb:05}.  Motivated by \citet{Yuan-Quat-Nara:03}, who
showed that the vertically-averaged electron density and temperature
are approximately power-laws in radius, we write the thermal electron
density, $n_e$, temperature $T_e$, and the non-thermal electron
density $n_{\rm nth}$ as
\begin{align}
n_e &= n^0_e \left(\frac{\rho}{M}\right)^{-1.1} \exp(-z^2/2\rho^2)
\nonumber\\
T_e &= T^0_e \left(\frac{r}{M}\right)^{-0.84} \label{model_eqs}
\\
n_{\rm nth} &= n^0_{\rm nth} \left(\frac{\rho}{M}\right)^{p_{\rm nth}}
\exp(-z^2/2\rho^2) \,,
\nonumber
\end{align}
where $\rho$ is the cylindrical radius relative to the black hole spin
axis, and $p_{\rm nth}$ is the radial power-law index for the
non-thermal electrons.  The assumed constants are listed in Table \ref{model_params}.
In all models, the non-thermal electrons have a spectral index of $1.25$
and a minimum Lorentz factor of 100.  
The magnetic field strength is set to be a fixed fraction ($30\%$) of
equipartition relative to the protons.  As suggested by recent
general-relativistic magnetohydrodynamic simulations, the field is taken to
be toroidal \citep{DeVi-Hawl-Krol:03}.  The accreting gas is assumed to be
in free fall inside of the ISCO ($6M$, $4.233M$ and $1.237M$ for $a=0$,
$0.5$ and $0.998$, respectively), and in Keplerian rotation otherwise.  In
all cases the disk angular momentum was aligned with the spin of the black
hole.

\subsection{Hot Spot Model} \label{CM:SM}
The hot spot is modeled by an overdensity of non-thermal electrons
centered at a point orbiting at the Keplerian velocity with a Gaussian
profile as measured in the comoving frame:
\begin{equation}
n_{\rm S} = n_{\rm S}^0
\exp\left[
-\frac{\Delta r^\mu \Delta r_\mu + \left( u_{\rm S}^\mu \Delta r_\mu\right)^2}{2 R_{\rm S}^2} \right]
\end{equation}
where $\Delta r^\mu \equiv r^\mu - r_{\rm S}^\mu$ is the
displacement from the spot centre (located at $r_{\rm S}^\mu$),
$u_{\rm S}^\mu$ is the spot four-velocity, $n_{\rm S}^0$ is the
spot central number density, and $R_{\rm S}$ is a measure of the spot
radius.  So that our results may be easily compared with
\citet{Brod-Loeb:05}, we used $R_{\rm S}=1.5 M$.
The spot central
number density was chosen to roughly reproduce the observed peak NIR flare flux.
As suggested by
coincident NIR--X-ray observations, the hot spot spectral index was
assumed to be $1.3$ \citep{Ecka_etal:04}, and the minimum Lorentz
factor was assumed to be $100$.  Such a situation may arise, \eg, in
the case of reconnection event similar to a solar flare.
\begin{figure}
\begin{center}
\includegraphics[width=\columnwidth]{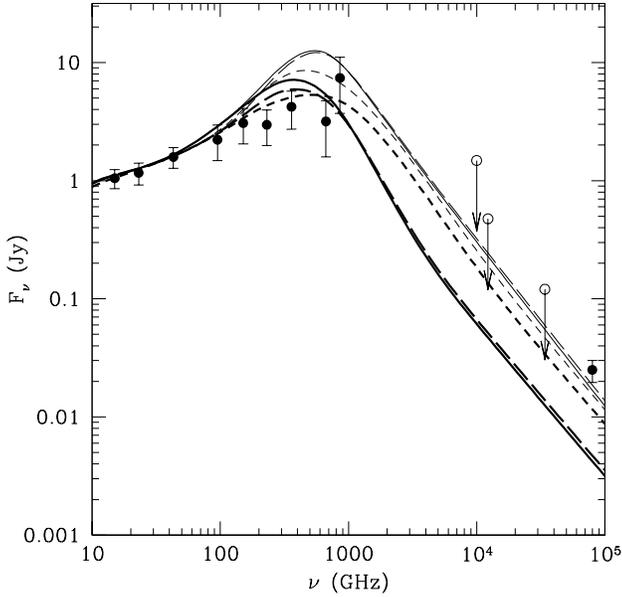}
\end{center}
\caption{The spectral flux for the non-rotating ($a=0$, solid),
  moderately rotating ($a=0.5$, long-dashed), and maximally rotating
  ($a=0.998$, short-dashed) \SgrA~disk models as viewed from
  $45^\circ$ degrees above the equatorial plane.  The thick lines
  correspond to the   quiescent disk emission and the thin lines
  correspond to the peak flare flux for a spot located at an orbital
  radius of $6M$.  The data points are taken from
  the compilation by \citet{Yuan-Quat-Nara:04}.}
\label{spectra_fig}
\end{figure}
The spectrum of the disk and our canonical flare model (at an orbital
radius of $6M$, viewed from $45^\circ$ above the equatorial plane) are
shown in Figure \ref{spectra_fig}.  Below $100\,\GHz$ the disk
photosphere extends beyond the hot spot, quenching its contribution to
the over all flux.  Above $100\,\GHz$ the hot spot becomes
increasingly visible, reaching a maximum near $500\,\GHz$.

For the Galactic centre, the orbital time scale near the ISCO are
between $5\,\min$ and $30\,\min$ for orbits around maximally and
non-rotating black holes, respectively.  The viability of a hot-spot
model for the flaring in \SgrA~requires that the synchrotron cooling
time scale be large in comparison to the orbital period.  Since the
emission at a frequency $\nu$ is dominated by electrons with Lorentz
factor $\gamma_e = \sqrt{\nu/\nu_B}$, where $\nu_B$ is the classical
electron cyclotron frequency, the cooling time of a flare as observed
at $\nu$ is
\begin{equation}
\tau_{\rm c}
\simeq
10^2 \left(\frac{\nu}{10^3\,\GHz}\right)^{-1/2} \left(\frac{B}{30 \G}\right)^{-3/2}\,\min\,.
\end{equation}
For the frequencies considered here this ranges from many hours (in
the sub-millimeter regime) to tens of minutes (in the NIR).  Hence, the
highest frequencies at which a substantial portion of the orbit is
likely to be visible are those in the NIR.

In order to concentrate upon the hot spot emission, all following plots
are of {\em background subtracted} quantities, \ie, the quiescent disk
emission has been removed.

\section{Near Infrared} \label{NIR}
As shown in \citet{Brod-Loeb:05b}, above $\sim 10^3\,\GHz$, the
accretion flow is optically thin everywhere.  As a result, at NIR
frequencies the emission from the hot spot is unadulterated.  For this
reason we discuss the light curves and centroid paths in the NIR
first.  Note that due to the fact that the spectra above $\sim
10^3\,\GHz$ has constant spectral index, the results of this section
apply to all wavelengths above the optically thick/thin transition.
These may be directly compared with the results of \citet{Brod-Loeb:05}.

\subsection{Light Curves} \label{NIR:LC}
\begin{figure}
\begin{center}
\includegraphics[width=\figfactor\columnwidth]{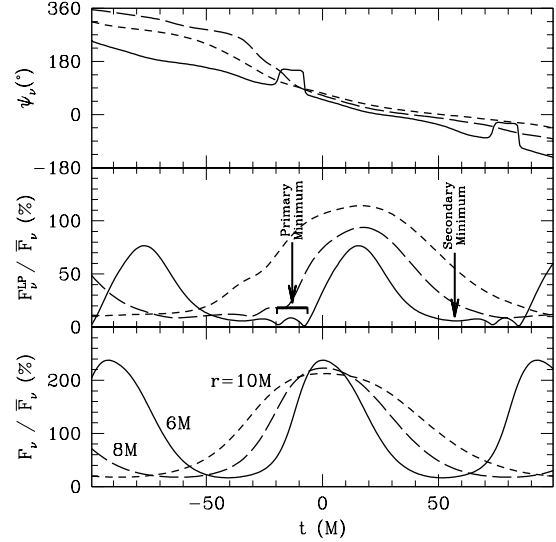}
\end{center}
\caption{The background subtracted total and polarized flux (bottom and
middle panels, respectively) normalized by the average {\em background
subtracted} total flux in the optically-thin regime as functions of time
for spot circular orbits at $6M$ (solid), $8M$ (long-dash) and $10M$
(short-dash) around a Schwarzschild black hole, viewed from $45^\circ$
above the equatorial plane.  The top panel shows the polarization angle,
$\psi$. The time axis is set so that a single orbital period of the $10M$
case is shown.  For a black hole mass of $4\times10^6\,\Msun$ (as in
\SgrA), the time unit is $M=20\,\s$.}
\label{lc_r_v}
\end{figure}
As in \citet{Brod-Loeb:05}, the primary distinction between hot spots
located at different radii is the time scale of variability in the
total flux.  This can be seen explicitly in the lower panel of Figure
\ref{lc_r_v} in which the magnification of the {\em hot spot emission}
is shown as a function of time.  This is evident in the polarized flux
variability as well (middle panel).  It should be noted that the
polarization variability will be different from that found in
\citet{Brod-Loeb:05} owing to the difference in the polarized emission
models (the one utilized here being the more physically motivated).
Nonetheless, the generic features of a primary minimum caused by the
development of the secondary hot-spot image (immediately preceding
maximum unpolarized magnification) and a weaker secondary minimum
caused by the development of a tertiary image (following the
maximum unpolarized magnification by $60M$) exist.  In this case the
polarization angle is punctuated by rapid rotations ($\sim 90^\circ$) at
the primary minimum, and uneventful otherwise, rotating as expected
for a toroidal field.

Characteristically, the peak polarized flux follows the peak
magnification.  In the canonical case of an spot orbital radius of $6M$
viewed from $45^\circ$ above the orbital plane, this corresponds to a
delay of approximately $5\,\min$ in the context of the Galactic
centre.  

\begin{figure}
\begin{center}
\includegraphics[width=\figfactor\columnwidth]{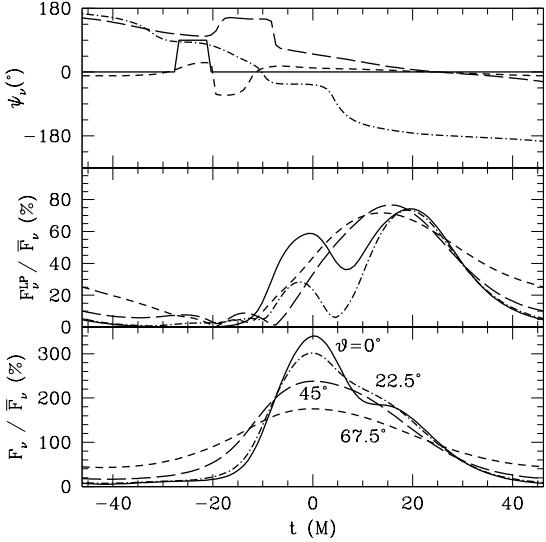}
\end{center}
\caption{The background subtracted total and polarized flux (bottom and
middle panels, respectively) normalized by the average {\em background
subtracted} total flux in the optically-thin regime as functions of time
for a spot orbit at $6M$ around a Schwarzschild black hole viewed from
$0^\circ$ (solid), $22.5^\circ$ (dash-dot), $45^\circ$ (long-dash) and
$67.5^\circ$ (short-dash) above the equatorial plane.  The polarization
angle, $\psi$, is shown in the top panel.  For a black hole mass of
$4\times10^6\,\Msun$ (as in \SgrA), the time unit is $M=20\,\s$.}
\label{lc_theta_v}
\end{figure}
Again, variations in viewing inclination, $\vartheta$, produces
similar effects as those reported in \citet{Brod-Loeb:05}, the
primary distinction being the total magnification.  There are
structural changes in the polarized flux, indicative of the prominence
of the primary and secondary hot-spot images.  Unlike
\citet{Brod-Loeb:05}, in the edge-on view ($\vartheta=0^\circ$),
strong gravitational lensing produces a polarization component
orthogonal to dominant polarization, leading to a rapid rotation in
polarization angle preceding peak magnification by approximately a
quarter-phase.

\begin{figure}
\begin{center}
\includegraphics[width=\figfactor\columnwidth]{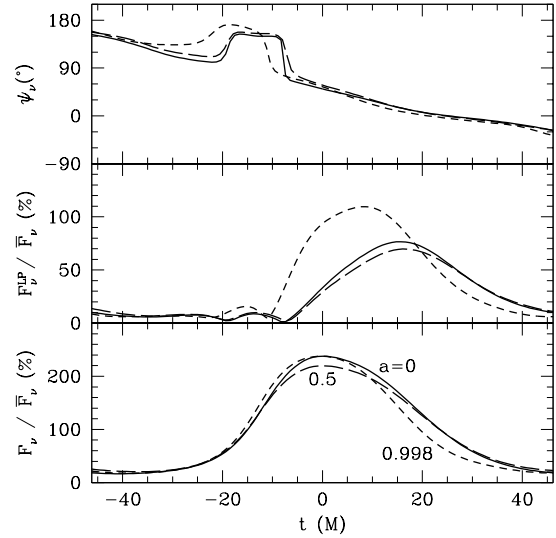}
\end{center}
\caption{The background subtracted total and polarized flux (bottom and
middle panels, respectively) normalized by the average {\em background
subtracted} total flux in the optically-thin regime as functions of time
for a spot orbit at $6M$ around a Kerr black hole with $a=0$ (solid),
$a=0.5$ (long-dash) and $a=0.998$ (short-dash) viewed from $45^\circ$ above
the equatorial plane.  The polarization angle, $\psi$, is shown in the top
panel.  The time axis is set so that a single orbital period of the $a=0$
case is shown.  For a black hole mass of $4\times10^6\,\Msun$ (as in
\SgrA), the time unit is $M=20\,\s$.}
\label{lc_a_v}
\end{figure}
The manifestation of black hole spin in the light curves is shown in Figure
\ref{lc_a_v}.  As in \citet{Brod-Loeb:05}, the differences among the
magnification light curves is small in comparison to those associated with
variations of the orbital parameters.  While the secondary minimum of the
polarization does vary significantly with spin, in the total flux this is
difficult to observe.

\begin{figure}
\begin{center}
\includegraphics[width=\figfactor\columnwidth]{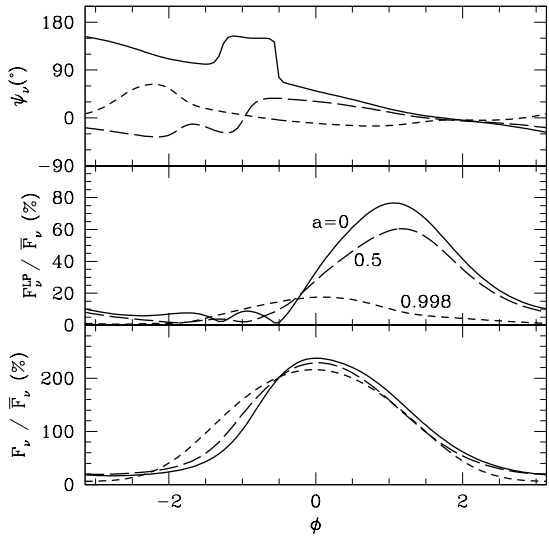}
\end{center}
\caption{The background subtracted total and polarized flux (bottom and
middle panels, respectively) normalized by the average {\em background
subtracted} total flux in the optically-thin regime as functions of phase
for spot orbits at the prograde ISCO around a Kerr black hole with $a=0$
(solid), $a=0.5$ (long-dash) and $a=0.998$ (short-dash) viewed from
$45^\circ$ above the equatorial plane.  The polarization angle, $\psi$, is
shown in the top panel.  The time axis is set so that a single orbital
period of the $a=0$ case is shown.  For a black hole mass of
$4\times10^6\,\Msun$ (as in \SgrA), the time unit is $M=20\,\s$.}
\label{lcph_ISCO_v}
\end{figure}
However, the primary distinction between different spins is likely to be
the variations in the radius of the ISCO, and thus the typical periods.
Figure \ref{lcph_ISCO_v} shows the magnification and polarization light
curves for a spot viewed from $45^\circ$ above the equatorial plane located
at the prograde ISCO for various dimensionless spin parameters (normalized
by $M$), namely $6M$, $4.233M$ and $1.237M$ for $a=0$, $0.5$ and $0.998$,
respectively.  Because the orbital timescales vary by nearly an order of
magnitude between $a=0$ and $a=0.998$, these are plotted as functions of
{\em orbital phase}.  Rapid rotation substantially reduces the maximum
polarized flux, primarily due to the enhanced gravitational lensing
associated with the compactness of the orbit.  However, as mentioned in
\citet{Brod-Loeb:05}, and alluded to above, the most significant
discriminator is likely to be the event time scale.

\subsection{Centroid Paths} \label{NIR:CP}
The PRIMA instrument at the VLT is expected to provide sub-milli-arcsecond
astrometry, thus enabling high resolution measurements of the the infrared
image centroid for \SgrA~and in turn constrain the accretion flow and black
hole parameters.  Because the location of the image centroids will be
dominated by the brightest features, subtracting the background accretion
flow emission is necessary to isolate the location of the hot spot during a
flare.  However, intrinsic variability in the source and/or the presence of
multiple hot spots may introduce substantial systematic errors into this
process.  Nonetheless, in order to clearly remove the uncertainty
associated with the accretion disk model, in computing all of the centroid
positions below we have used the {\em background subtracted} images.  These
may be generated using the centroids of the total emission if the quiescent
centroid position and intensity are known:
\begin{equation}
\overline{\bmath{X}}_{\rm S}(t)
=
\frac{F(t)\overline{\bmath{X}}(t)-F_{\rm BG}\overline{\bmath{X}}_{\rm BG}}
{ F(t) - F_{\rm BG} }\,,
\end{equation}
where $\overline{\bmath{X}}_{\rm S}(t)$ is the location of the spot
centroid, $\overline{\bmath{X}}(t)$ and $F(t)$ are the location of the
total image centroid and the total observed flux, and
$\overline{\bmath{X}}_{\rm BG}$ and $F_{\rm BG}$ are the background
(disk) image centroid and flux.

\begin{figure}
\begin{center}
\includegraphics[width=\figfactor\columnwidth]{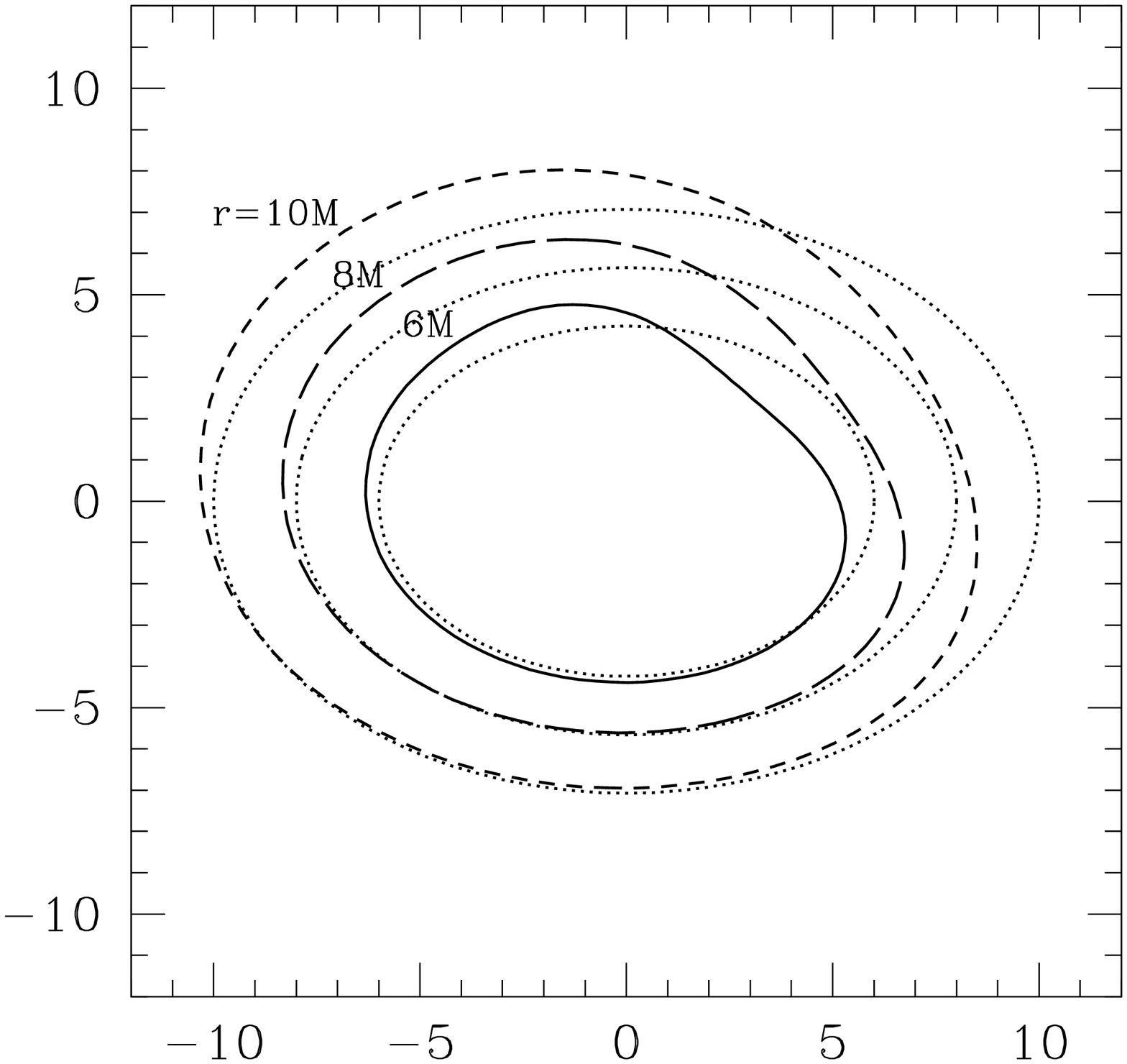}
\end{center}
\caption{The paths of the background subtracted intensity centroid in the
optically-thin regime for circular spot orbits around a Schwarzschild black
hole viewed from $45^\circ$ above the orbital plane with radii $6M$
(solid), $8M$ (long-dash) and $10M$ (short dash).  For reference, a circle
inclined at $45^\circ$ is also shown by the dotted lines for each radius.
Axes are labeled in units of $M$ (corresponding to an angular scale of
roughly $5\,\muas$ for \SgrA).}
\label{cents_r_v}
\end{figure}
\begin{figure}
\begin{center}
\includegraphics[width=\figfactor\columnwidth]{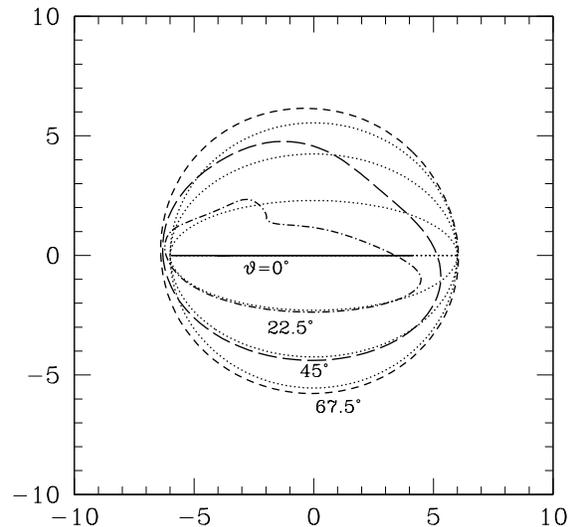}
\end{center}
\caption{The paths of the background subtracted intensity centroid in the
optically-thin regime for circular spot orbits around a Schwarzschild black
hole with radius $6M$ viewed from $0^\circ$ (solid), $22.5^\circ$
(dash-dot), $45^\circ$ (long-dash) and $67.5^\circ$ (short-dash) above the
orbital plane.  For reference, a circles inclined at each angle are shown
by the dotted lines. Axes are labeled in units of $M$ (corresponding to an
angular scale of roughly $5\,\muas$ for \SgrA).}
\label{cents_theta_v}
\end{figure}
While gravitational lensing and relativistic aberration distort the
centroid path, it is nevertheless possible to determine the orbital radius
(Figure \ref{cents_r_v}) and inclination (Figure \ref{cents_theta_v}).
Combined with the orbital period at an orbital radius $r_S$ (in units of M),
\begin{equation}
P = 2\pi M \left( r_S^{3/2}-a \right)\,,
\end{equation}
(where positive $a$ corresponds to prograde orbits) determined, \eg, by the
magnification light curve, this provides a direct measurement of the spin.

\begin{figure}
\begin{center}
\includegraphics[width=\figfactor\columnwidth]{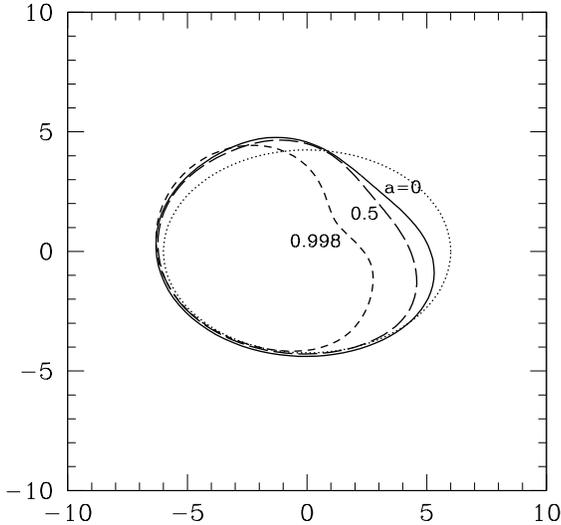}
\end{center}
\caption{The paths of the background subtracted intensity centroid in the
optically-thin regime for circular spot orbits with radius $6M$ in the
equatorial plane around a Kerr black hole viewed from $45^\circ$ above the
orbital plane for $a=0$ (solid), $a=0.5$ (long-dash) and $a=0.998$
(short-dash).  For reference, a circle inclined at $45^\circ$ is also shown
by the dotted line.  Axes are labeled in units of $M$ (corresponding to an
angular scale of roughly $5\,\muas$ for \SgrA).}
\label{cents_a_v}
\end{figure}
\begin{figure}
\begin{center}
\includegraphics[width=\figfactor\columnwidth]{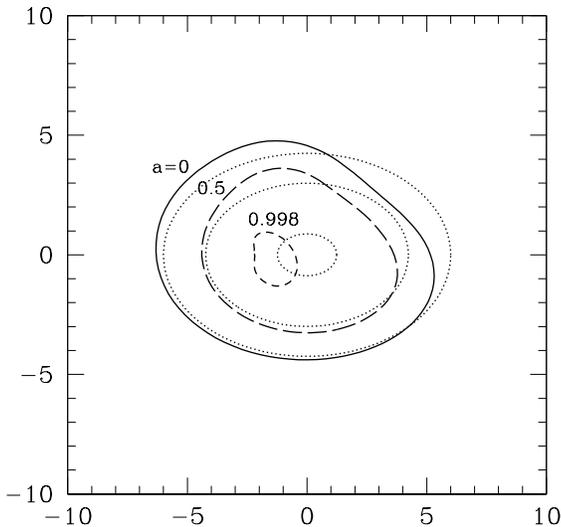}
\end{center}
\caption{The paths of the background subtracted intensity centroid in the
optically-thin regime for circular spot orbits around a Kerr black hole
viewed from $45^\circ$ above the orbital plane at the prograde ISCO for
$a=0$ (solid), $a=0.5$ (long-dash) and $a=0.998$ (short-dash).  For
reference, a circle inclined at $45^\circ$ is also shown by the dotted
lines for spin.  Axes are labeled in units of $M$ (corresponding to an
angular scale of roughly $5\,\muas$ for \SgrA).}
\label{cents_ISCO_v}
\end{figure}
Figures \ref{cents_a_v} and \ref{cents_ISCO_v} present the effects
of black hole spin upon the centroid paths.  In Figure \ref{cents_a_v}
the hot spot is located at $6M$ and viewed from $45^\circ$ above the
orbital plane.  In this case a moderate increase in the orbital
velocity with spin leads to an enhancement of the relativistic
beaming, shifting the centroid path towards the approaching side
(left).  In Figure \ref{cents_ISCO_v} the hot spot is located at the
prograde ISCO for each black hole spin, again viewed from $45^\circ$
above the orbital plane.  Here the centroid path at high spin, where
the ISCO is close to the horizon, is substantially offset.  However,
in the case of moderate spin, the generic dependence upon radius
seen in Figure \ref{cents_r_v} is dominant.

\section{Generic Effects of Opacity} \label{GEoO}
While in the NIR the disk opacity to synchrotron self-absorption is
negligible, this is not the case at sub-millimeter wavelengths, at which
the accretion flow is only beginning to be optically thin.  As discussed in
\citet{Brod-Loeb:05b}, the accretion flow opacity is not symmetric between
the two sides of the disk, and is enhanced by the Doppler effect on the
approaching side of the disk.  As a result, the light curves and centroid
paths presented in the previous section for the NIR can be substantially
modified for frequencies near the optically thick/thin transition.  For the
purpose of highlighting the generic effects of opacity, and the utility of
high-resolution multi-wavelength flare observations, we present next a
comparison of the light curves and centroid paths for NIR and radio
frequencies.

\subsection{Time Averaged Images}
\begin{figure*}
\begin{center}
\includegraphics[width=\textwidth]{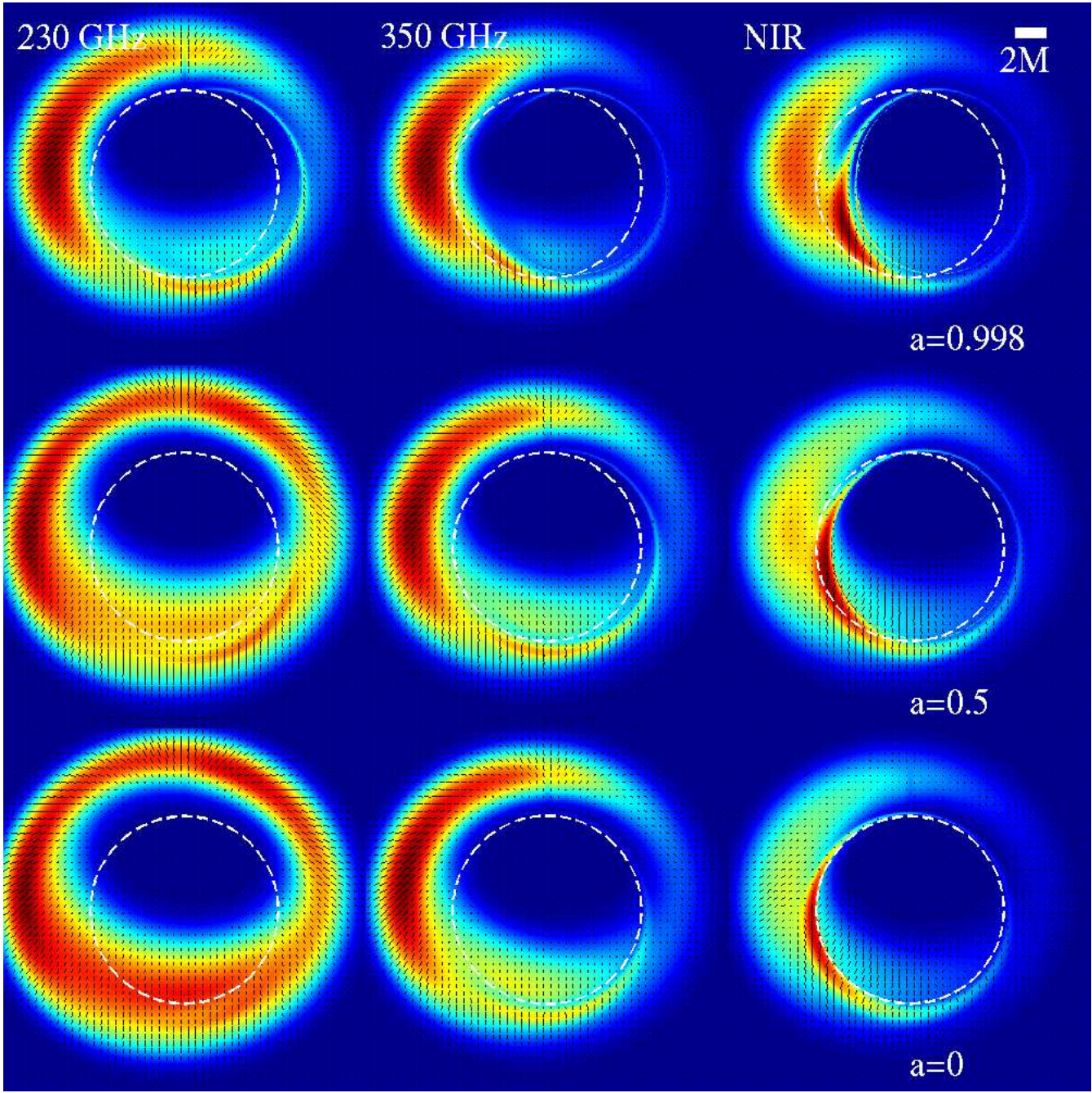}
\end{center}
\caption{Comparison of the orbit-averaged, disk-subtracted images of a spot
for two radio frequencies (at which opacity is important) and in the
infrared (at which the disk is transparent).  For reference, the
photon-capture cross section for a Schwarzschild black hole is shown by the
dashed white line.}
\label{spot_avgs}
\end{figure*}
The effects of asymmetric opacity are readily apparent in the
orbit-averaged images, shown in Figure \ref{spot_avgs} for the
canonical hot-spot model for the three disk models discussed in
Section \ref{CM:DM}, as observed at $230\,\GHz$, $350\,\GHz$ and in the
NIR.  In the NIR both the direct and secondary images are clearly visible,
corresponding to the larger and smaller rings, respectively.  Less
visible are the tertiary images, which produce the very thin ring
directly inside that associated with the secondary images.  In
addition, as expected from the Doppler shift and relativistic beaming,
the approaching portion of the hot-spot orbit (left side) appears
considerably brighter than the receding portion of the spot orbit
(right side).  At radio frequencies the secondary image is
substantially less visible and the brightness contrast between the
approaching and receding portions of the hot-spot orbit decreases with
decreasing frequency.  This is a direct result of the increased
opacity in the approaching portion of the disk (also on the left
side).  As mentioned in \citet{Brod-Loeb:05}, in the limit of a
thermal spectrum the orbit-averaged images are symmetric, and thus it
is not surprising that increased opacity results in more symmetric
images.

Also shown in Figure \ref{spot_avgs} is the degree
and orientation of the polarized flux.  In all cases this is dominated
by the approaching portion of the orbit.  In the NIR, a second
polarization component is present due to the secondary image.  This
is substantially suppressed at radio wavelengths, and thus the strong
swings in polarization angle are not expected in the radio.  Finally,
the degree of polarization is also expected to decrease as a result of
the decrease in the brightness asymmetry mentioned before.

\subsection{Light Curves}
\begin{figure}
\begin{center}
\includegraphics[width=\figfactor\columnwidth]{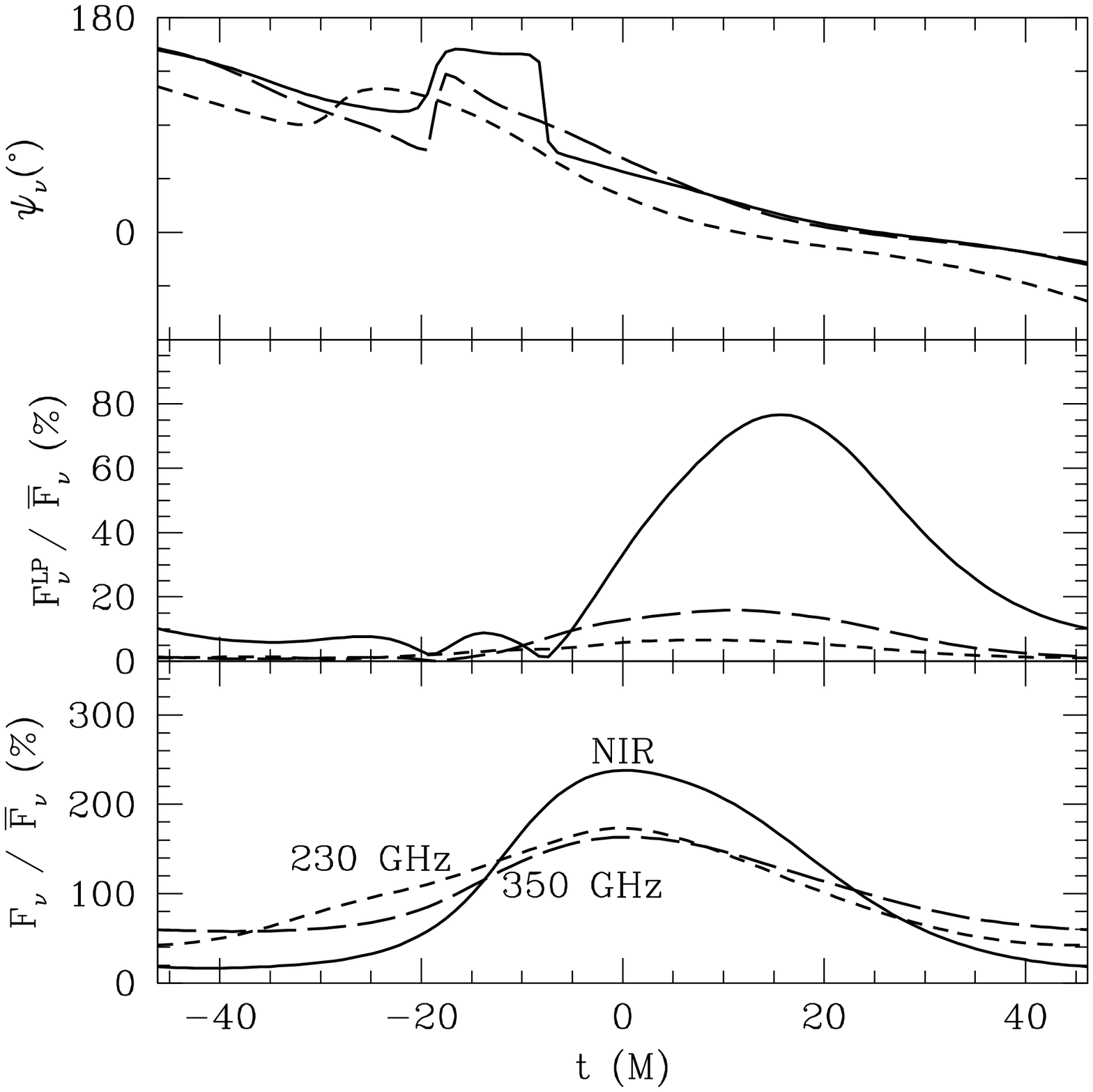}
\end{center}
\caption{The background subtracted total and polarized flux (bottom and
middle panels, respectively) normalized by the average {\em background
subtracted} total flux as functions of time for a spot orbit at $6M$ around
a Schwarzschild black hole viewed from $45^\circ$ above the equatorial
plane as observed in the infrared (solid), $350\,\GHz$ (long-dash) and
$230\,\GHz$ (short-dash).  The polarization angle, $\psi$, is shown in the
top panel.  The time axis is set so that a single orbital period of the
$a=0$ case is shown.  For a black hole mass of $4\times10^6\,\Msun$ (as in
\SgrA), the time unit is $M=20\,\s$.}
\label{lc_nu_a0}
\end{figure}
\begin{figure}
\begin{center}
\includegraphics[width=\figfactor\columnwidth]{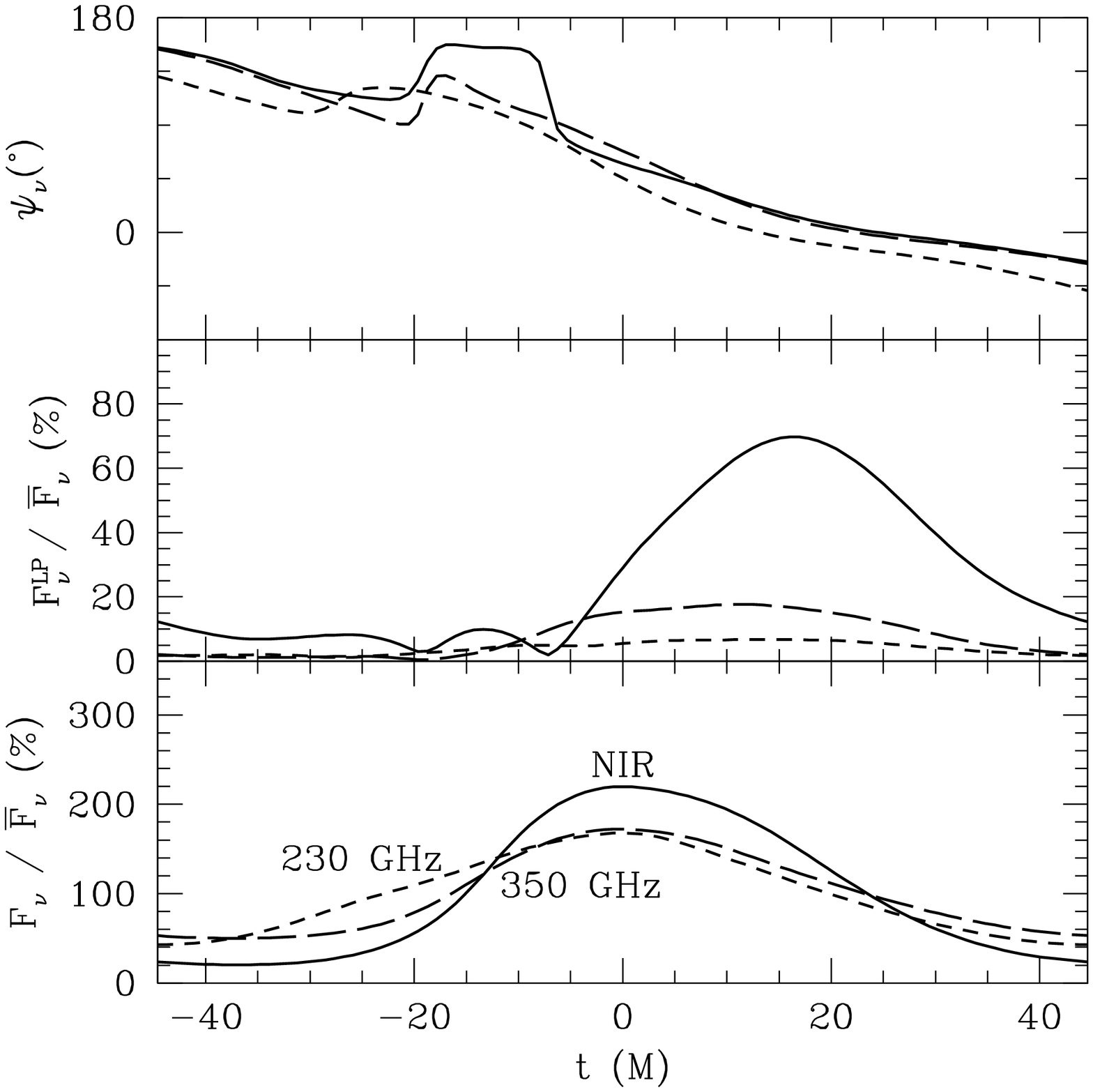}
\end{center}
\caption{The background subtracted total and polarized flux (bottom and
middle panels, respectively) normalized by the average {\em background
subtracted} total flux as functions of time for a spot orbit at $6M$ around
a Kerr black hole ($a=0.5$) viewed from $45^\circ$ above the equatorial
plane as observed in the infrared (solid), $350\,\GHz$ (long-dash) and
$230\,\GHz$ (short-dash).  The polarization angle, $\psi$, is shown in the
top panel.  The time axis is set so that a single orbital period of the
$a=0$ case is shown.  For a black hole mass of $4\times10^6\,\Msun$ (as in
\SgrA), the time unit is $M=20\,\s$.}
\label{lc_nu_a5}
\end{figure}
\begin{figure}
\begin{center}
\includegraphics[width=\figfactor\columnwidth]{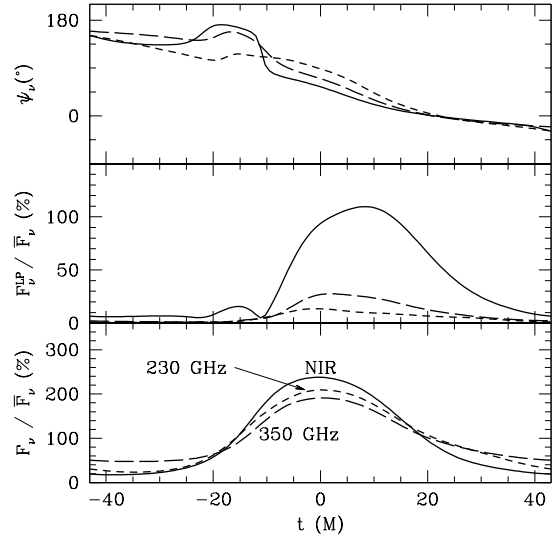}
\end{center}
\caption{The background subtracted total and polarized flux (bottom and
middle panels, respectively) normalized by the average {\em background
subtracted} total flux as functions of time for a spot orbit at $6M$ around
a Kerr black hole ($a=0.998$) viewed from $45^\circ$ above the equatorial
plane as observed in the infrared (solid), $350\,\GHz$ (long-dash) and
$230\,\GHz$ (short-dash).  The polarization angle, $\psi$, is shown in the
top panel.  The time axis is set so that a single orbital period of the
$a=0$ case is shown.  For a black hole mass of $4\times10^6\,\Msun$ (as in
\SgrA), the time unit is $M=20\,\s$.}
\label{lc_nu_a998}
\end{figure}
Figures \ref{lc_nu_a0}--\ref{lc_nu_a998} shown the magnification and
polarized flux light curves in the radio and NIR for a hot-spot
orbiting an $a=0$, $0.5$ and $0.998$ black hole, respectively.
The primary distinction between the different observing wavelengths is
in the degree of polarization, decreasing by roughly an order of
magnitude from the NIR to $230\,\GHz$.  In addition, as expected from
the images, the polarization angle light curve is progressively
smoothed as the observing wavelength increases.  It should also be
noted that the secondary minimum in the polarization, also resulting
from the development of tertiary images, is significantly less
visible in the radio.

\subsection{Centroid Paths}

\begin{figure}
\begin{center}
\includegraphics[width=\figfactor\columnwidth]{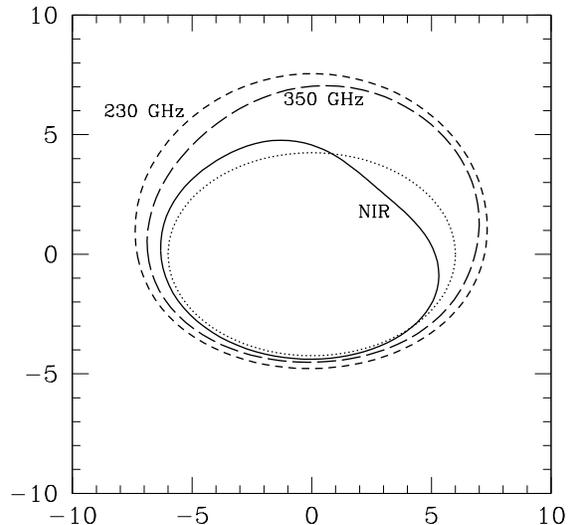}
\end{center}
\caption{The paths of the background subtracted intensity centroid for
circular spot orbits around a Schwarzschild black hole viewed from
$45^\circ$ above the orbital plane at $6M$ as observed in the infrared
(solid), $350\,\GHz$ (long-dash) and $230\,\GHz$ (short-dash).  For
reference, a circle inclined at $45^\circ$ is also shown by the dotted
lines for each radius.  Axes are labeled in units of $M$ (corresponding to
an angular scale of roughly $5\,\muas$ for \SgrA).}
\label{cents_nu_a0}
\end{figure}
\begin{figure}
\begin{center}
\includegraphics[width=\figfactor\columnwidth]{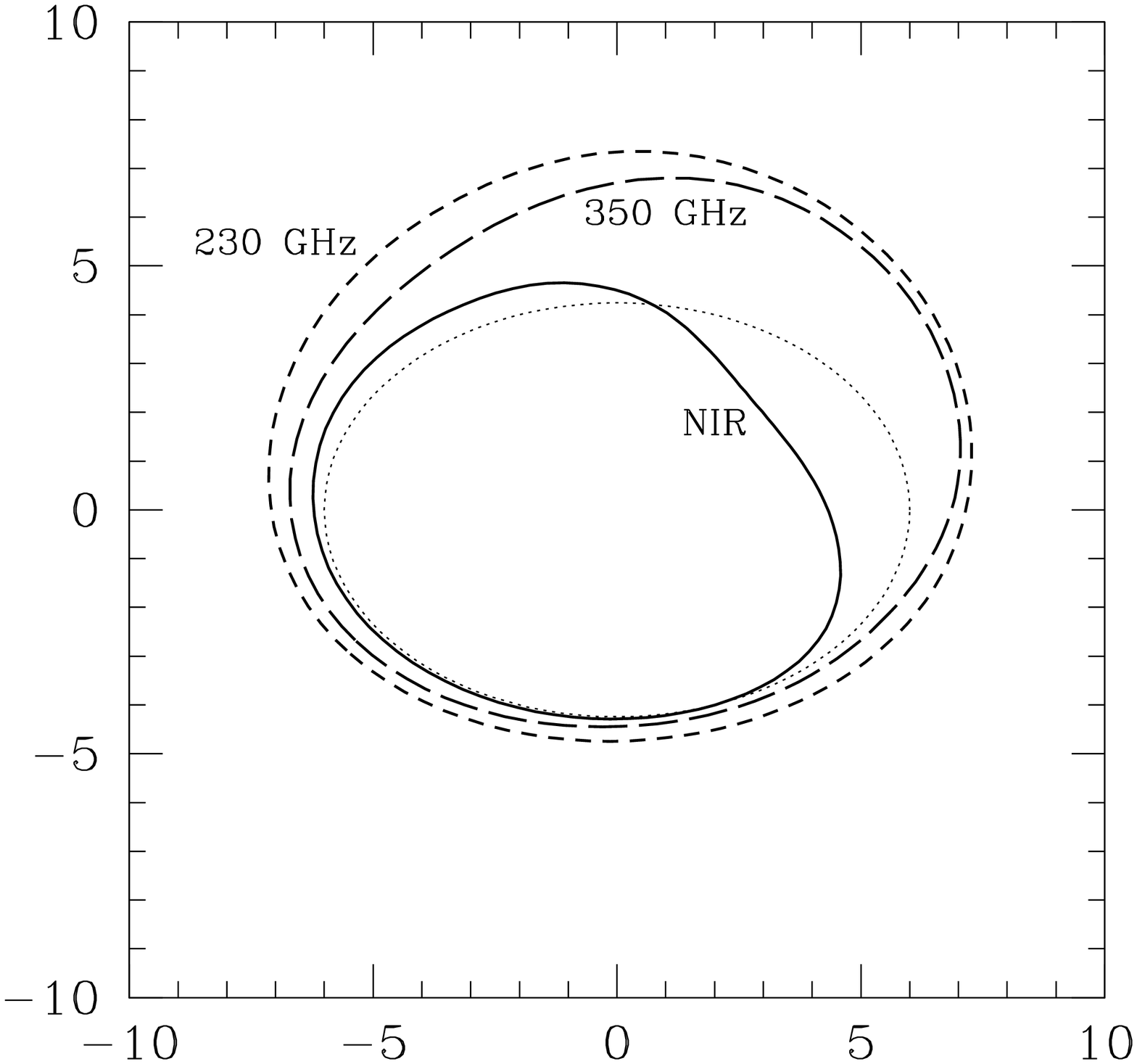}
\end{center}
\caption{The paths of the background subtracted intensity centroid for
circular spot orbits around a Kerr black hole ($a=0.5$) viewed from
$45^\circ$ above the orbital plane at $6M$ as observed in the infrared
(solid), $350\,\GHz$ (long-dash) and $230\,\GHz$ (short-dash).  For
reference, a circle inclined at $45^\circ$ is also shown by the dotted
lines for each radius.  Axes are labeled in units of $M$ (corresponding to
an angular scale of roughly $5\,\muas$ for \SgrA).}
\label{cents_nu_a5}
\end{figure}
\begin{figure}
\begin{center}
\includegraphics[width=\figfactor\columnwidth]{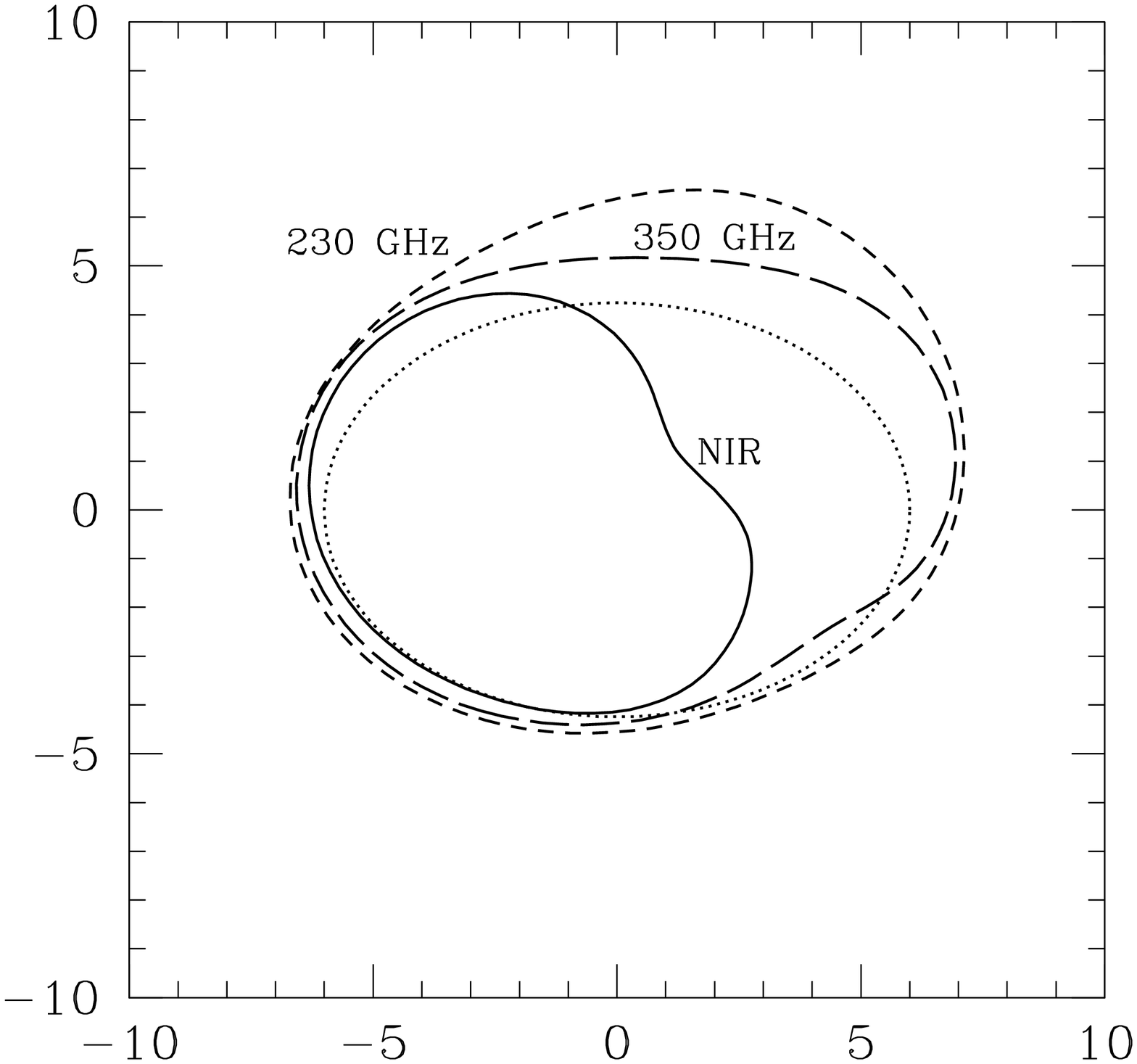}
\end{center}
\caption{The paths of the background subtracted intensity centroid for
circular spot orbits around a Kerr black hole ($a=0.998$) viewed from
$45^\circ$ above the orbital plane at $6M$ as observed in the infrared
(solid), $350\,\GHz$ (long-dash) and $230\,\GHz$ (short-dash).  For
reference, a circle inclined at $45^\circ$ is also shown by the dotted
lines for each radius.  Axes are labeled in units of $M$ (corresponding to
an angular scale of roughly $5\,\muas$ for \SgrA).}
\label{cents_nu_a998}
\end{figure}
The centroid paths for different observing frequencies are compared in
Figures \ref{cents_nu_a0}--\ref{cents_nu_a998}.  As the observing frequency
decreases the centroid path becomes larger and more symmetric.  As a
result, the orbital parameters are more easily determined at radio
wavelengths.  This implies that simultaneous high-resolution radio and NIR
observations of the centroid are capable of identifying black hole spin.

\section{Radio} \label{R}
Anticipating high-resolution sub-millimetre observations of the
Galactic centre, we present the magnification and polarized flux light
curves and image centroids at radio frequencies.  As discussed in the
previous section, these may be expected to be quantitatively different
from those presented in Section \ref{NIR}.

\subsection{Light Curves} \label{R:LC}

\begin{figure}
\begin{center}
\includegraphics[width=\figfactor\columnwidth]{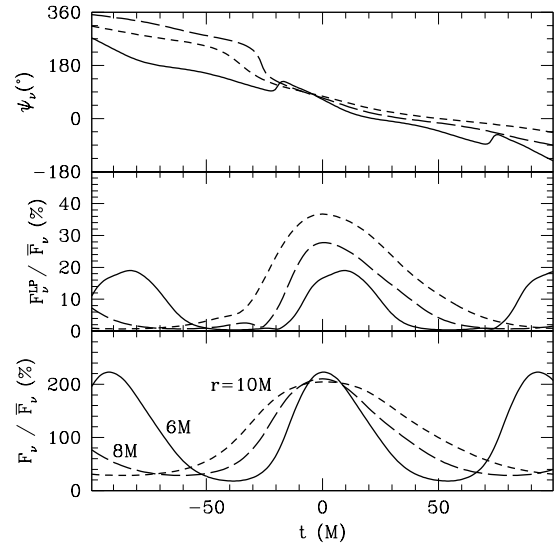}
\end{center}
\caption{The background subtracted total and polarized flux (bottom and
middle panels, respectively) normalized by the average {\em background
subtracted} total flux as functions of time at $350\,\GHz$ for spot orbits
at $6M$ (solid), $8M$ (long-dash) and $10M$ (short-dash) around a
Schwarzschild black hole, viewed from $45^\circ$ above the equatorial
plane.  The polarization angle, $\psi$, is shown in the top panel.  The
time axis is set so that a single orbital period of the $10M$ case is
shown.  For a black hole mass of $4\times10^6\,\Msun$ (as in \SgrA), the
time unit is $M=20\,\s$.}
\label{lc_r_350}
\end{figure}
\begin{figure}
\begin{center}
\includegraphics[width=\figfactor\columnwidth]{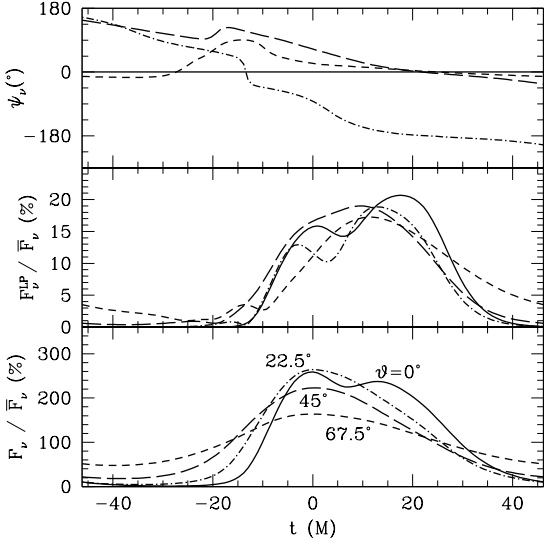}
\end{center}
\caption{The background subtracted total and polarized flux (bottom and
middle panels, respectively) normalized by the average {\em background
subtracted} total flux as functions of time at $350\,\GHz$ for a spot orbit
at $6M$ around a Schwarzschild black hole viewed from $0^\circ$ (solid),
$22.5^\circ$ (dash-dot), $45^\circ$ (long-dash) and $67.5^\circ$
(short-dash) above the equatorial plane.  The polarization angle, $\psi$,
is shown in the top panel.  For a black hole mass of $4\times10^6\,\Msun$
(as in \SgrA), the time unit is $M=20\,\s$.}
\label{lc_theta_350}
\end{figure}
Figure \ref{lc_r_350} shows that at $350\,\GHz$ the absorption has little
effect upon the magnification light curves for moderate orbital
inclinations (compare to Figure \ref{lc_r_v}).  However, the peak polarized
flux is reduced by more than $70\%$ and observed earlier than in the
NIR. While the polarization angle continues to exhibit a rapid rotation
associated with the development of a secondary image, it is significantly
smoothed.

This is especially apparent in Figure \ref{lc_theta_350}, which shows the
light curves for a number of viewing inclinations, $\vartheta$, (compare to
Figure \ref{lc_theta_v}).  In the case where $\vartheta=0^\circ$, the
emergence of an orthogonal polarization is suppressed, leading to a
constant polarization angle.  For small $\vartheta$, the peak magnification
is also marginally reduced, as may be expected from the larger column
density for such viewing angles.  Nevertheless, the flare time-scale and
peak magnification are still indicative of the orbital radius and
inclination.  Furthermore, the presence of a flare in the polarized flux is
generic.

\begin{figure}
\begin{center}
\includegraphics[width=\figfactor\columnwidth]{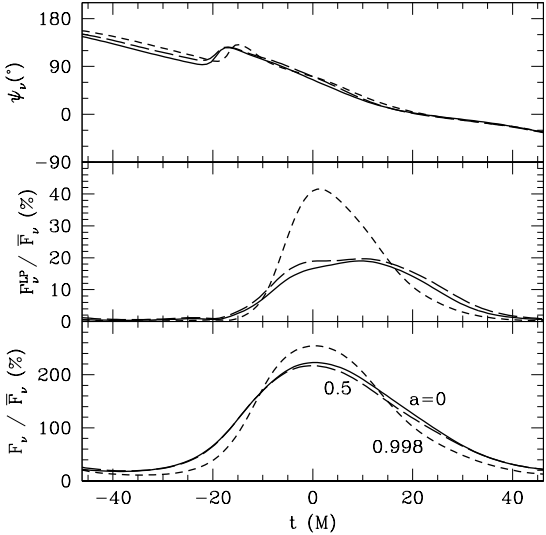}
\end{center}
\caption{The background subtracted total and polarized flux (bottom and
middle panels, respectively) normalized by the average {\em background
subtracted} total flux as functions of time at $350\,\GHz$ for a spot orbit
at $6M$ around a Kerr black hole with $a=0$ (solid), $a=0.5$ (long-dash)
and $a=0.998$ (short-dash) viewed from $45^\circ$ above the equatorial
plane.  The polarization angle, $\psi$, is shown in the top panel.  The
time axis is set so that a single orbital period of the $a=0$ case is
shown.  For a black hole mass of $4\times10^6\,\Msun$ (as in \SgrA), the
time unit is $M=20\,\s$.}
\label{lc_a_350}
\end{figure}
\begin{figure}
\begin{center}
\includegraphics[width=\figfactor\columnwidth]{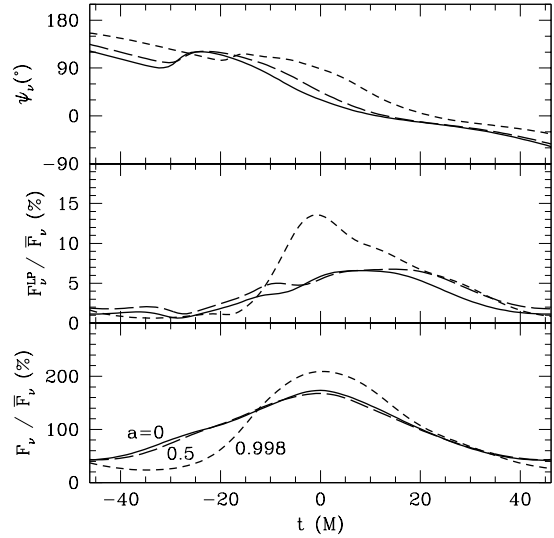}
\end{center}
\caption{The background subtracted total and polarized flux (bottom and
middle panels, respectively) normalized by the average {\em background
subtracted} total flux as functions of time at $230\,\GHz$ for a spot orbit
at $6M$ around a Kerr black hole with $a=0$ (solid), $a=0.5$ (long-dash)
and $a=0.998$ (short-dash) viewed from $45^\circ$ above the equatorial
plane.  The polarization angle, $\psi$, is shown in the top panel.  The
time axis is set so that a single orbital period of the $a=0$ case is
shown.  For a black hole mass of $4\times10^6\,\Msun$ (as in \SgrA), the
time unit is $M=20\,\s$.}
\label{lc_a_230}
\end{figure}
Figures \ref{lc_a_350} and \ref{lc_a_230} show the magnification and
polarized flux light curves for various black hole spins as seen at
$350\,\GHz$ and $230\,\GHz$, respectively (compare to Figure \ref{lc_a_v}).
Typically, there is little difference among the magnification light curves.
However, in both cases the polarized flux clearly discriminates between
high and moderate/low black hole spins, varying by roughly a factor of two
in both cases.  This is likely due to the lower thermal electron density
permissible in the high spin model employed (see Table \ref{model_params}).

\begin{figure}
\begin{center}
\includegraphics[width=\figfactor\columnwidth]{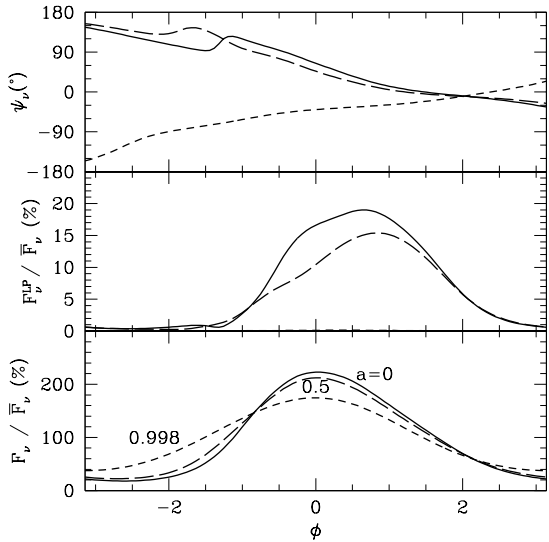}
\end{center}
\caption{The background subtracted total and polarized flux (bottom and
middle panels, respectively) normalized by the average {\em background
subtracted} total flux as functions of phase at $350\,\GHz$ for spot orbits
at the prograde ISCO around a Kerr black hole with $a=0$ (solid), $a=0.5$
(long-dash) and $a=0.998$ (short-dash) viewed from $45^\circ$ above the
equatorial plane.  The polarization angle, $\psi$, is shown in the top
panel.  The time axis is set so that a single orbital period of the $a=0$
case is shown.  For a black hole mass of $4\times10^6\,\Msun$ (as in
\SgrA), the time unit is $M=20\,\s$.}
\label{lcph_ISCO_350}
\end{figure}
As in the optically-thin regime described in Section \ref{NIR:LC}, the
primary difference between high and low spin black holes is likely to be
the flaring timescales.  Similarly to Figure \ref{lcph_ISCO_v}, Figure
\ref{lcph_ISCO_350} shows the magnification and polarized flux light curves
as functions of {\em orbital phase}.  As in the NIR, the polarization is
more sensitive to the black hole spin than the degree of magnification.
The opacity associated with the hot-spot orbiting the maximally rotating
black hole is significantly larger due primarily to the rapid disk velocity
at this position.  As a result, this model was significantly more absorbed
than the others.  Combined with strong gravitational lensing, this effect
produces a variation in the flux that is unpolarized and small in
comparison with the quiescent disk emission.

\subsection{Centroid Paths} \label{R:CP}
Next, we consider the paths of the {\em background subtracted} image
centroids.  While in the optically thin regime described in Section
\ref{NIR:CP}, the disk and hot-spot emission are completely disentangled,
in this case the disk opacity can substantially modify the image centroids.
In the frequency regime of the optically thick/thin transition, opacity is
most significant on the approaching side of the accretion flow, near where
the hot-spot is the brightest.  This has the effect of reducing the
brightness contrast between the approaching and receding portions of the
hot-spot orbit.  Since the brightest regions dominate the location of the
image centroid, this has significant implications for the centroid paths.

\begin{figure}
\begin{center}
\includegraphics[width=\figfactor\columnwidth]{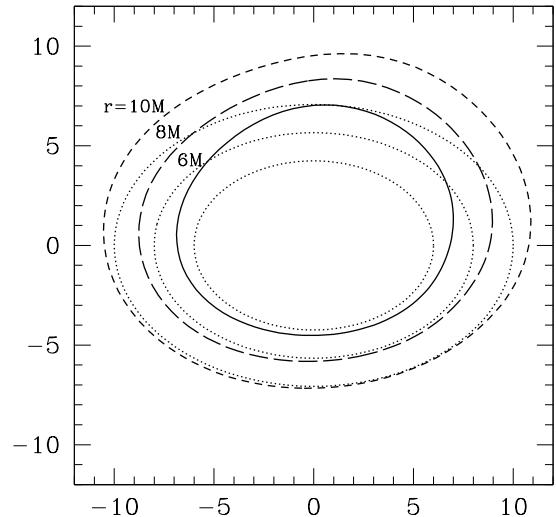}
\end{center}
\caption{The paths of the background subtracted intensity centroid at
$350\,\GHz$ for circular spot orbits around a Schwarzschild black hole
viewed from $45^\circ$ above the orbital plane with radii $6M$ (solid),
$8M$ (long-dash) and $10M$ (short dash).  For reference, a circle inclined
at $45^\circ$ is also shown by the dotted lines for each radius.  Axes are
labeled in units of $M$ (corresponding to an angular scale of roughly
$5\,\muas$ for \SgrA).}
\label{cents_r_350}
\end{figure}
Figure \ref{cents_r_350} shows the centroid paths for hot-spots orbiting at
three radii, viewed from $45^\circ$ above the orbital plane.  This is
similar to Figure 12 in \citet{Brod-Loeb:05},
primarily due to the reduced brightness contrast
mentioned above.  The orbits are noticeably larger than those in
Figure \ref{cents_r_v}, though the major axis distance is similar.

\begin{figure}
\begin{center}
\includegraphics[width=\figfactor\columnwidth]{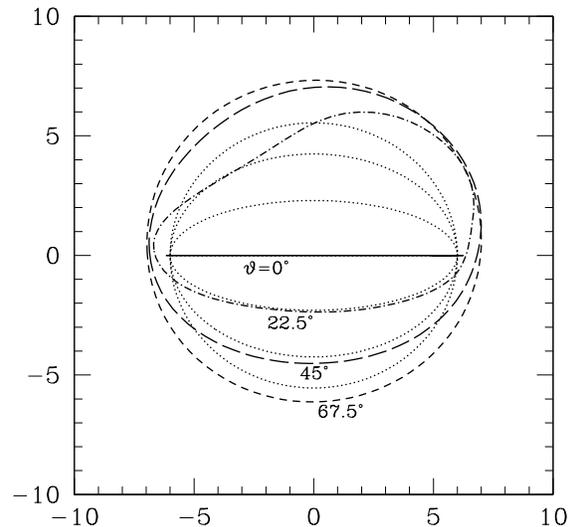}
\end{center}
\caption{The paths of the background subtracted intensity centroid at
$350\,\GHz$ for circular spot orbits around a Schwarzschild black hole with
radius $6M$ viewed from $0^\circ$ (solid), $22.5^\circ$ (dash-dot),
$45^\circ$ (long-dash) and $67.5^\circ$ (short-dash) above the orbital
plane.  For reference, a circles inclined at each angle are shown by the
dotted lines. Axes are labeled in units of $M$ (corresponding to an angular
scale of roughly $5\,\muas$ for \SgrA).}
\label{cents_theta_350}
\end{figure}
Various viewing inclinations are shown in Figure \ref{cents_theta_350}
(\cf, Figure \ref{cents_theta_v}).  As in Figure \ref{cents_r_350}, the
centroid paths are larger than those found in the NIR.  For low
$\vartheta$, this substantially modifies the shape of the centroid path.
Nonetheless, the nearest region of the orbit (bottom) is close to the
unlensed orbit positions (shown by the dotted lines), and thus the
semi-minor axis is still diagnostic of the orbital inclination.

\begin{figure}
\begin{center}
\includegraphics[width=\figfactor\columnwidth]{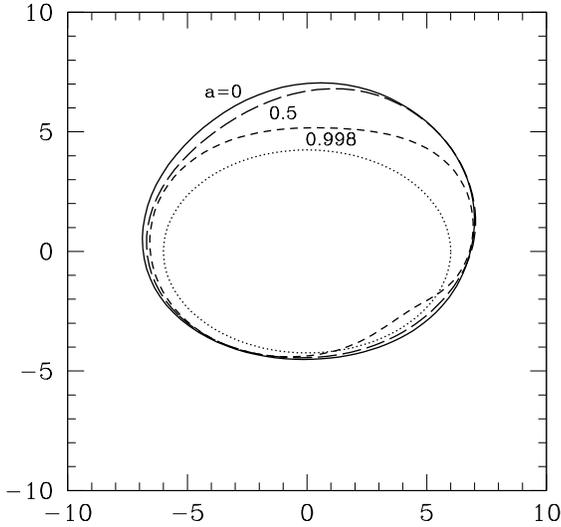}
\end{center}
\caption{The paths of the background subtracted intensity centroid at
$350\,\GHz$ for circular spot orbits with radius $6M$ in the equatorial
plane around a Kerr black hole viewed from $45^\circ$ above the orbital
plane for $a=0$ (solid), $a=0.5$ (long-dash) and $a=0.998$ (short-dash).
For reference, a circle inclined at $45^\circ$ is also shown by the dotted
line.  Axes are labeled in units of $M$ (corresponding to an angular scale
of roughly $5\,\muas$ for \SgrA).}
\label{cents_a_350}
\end{figure}
\begin{figure}
\begin{center}
\includegraphics[width=\figfactor\columnwidth]{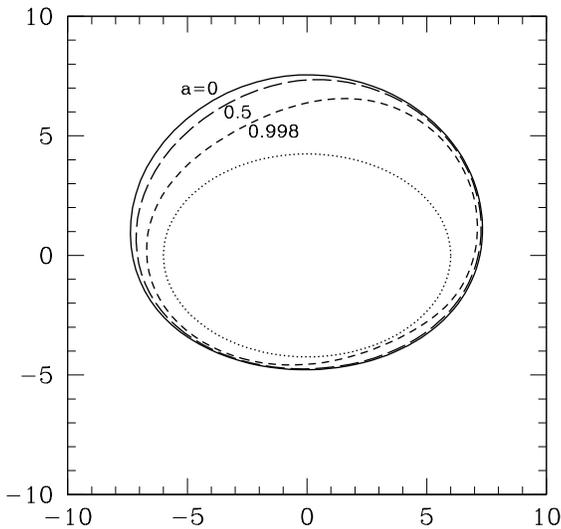}
\end{center}
\caption{The paths of the background subtracted intensity centroid at
$230\,\GHz$ for circular spot orbits with radius $6M$ in the equatorial
plane around a Kerr black hole viewed from $45^\circ$ above the orbital
plane for $a=0$ (solid), $a=0.5$ (long-dash) and $a=0.998$ (short-dash).
For reference, a circle inclined at $45^\circ$ is also shown by the dotted
line.  Axes are labeled in units of $M$ (corresponding to an angular scale
of roughly $5\,\muas$ for \SgrA).}
\label{cents_a_230}
\end{figure}
Figures \ref{cents_a_350} and \ref{cents_a_230} show the centroid
paths for various black hole spins as viewed at $350\,\GHz$ and
$230\,\GHz$, respectively (\cf, Figure \ref{cents_a_v}).  In all cases
the paths are larger than in the NIR and, unlike the NIR, are larger
than expected in the absence of gravitational and opacity effects
(shown by the dotted line).  In all cases, the semi-minor axis again
appears to be a good measure of the orbital inclination.

\begin{figure}
\begin{center}
\includegraphics[width=\figfactor\columnwidth]{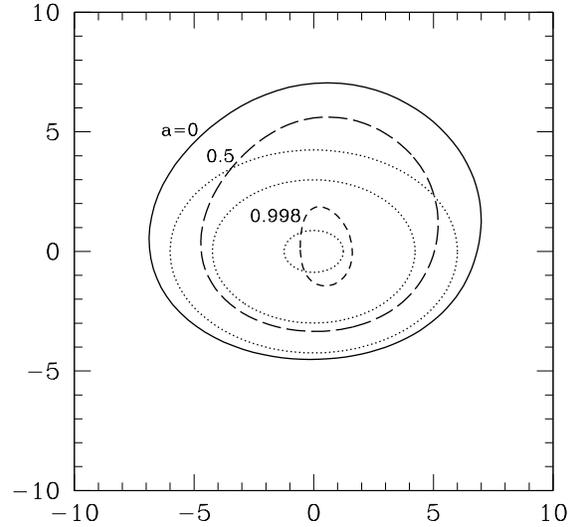}
\end{center}
\caption{The paths of the background subtracted intensity centroid at
$350\,\GHz$ for circular spot orbits around a Kerr black hole viewed from
$45^\circ$ above the orbital plane at the prograde ISCO for $a=0$ (solid),
$a=0.5$ (long-dash) and $a=0.998$ (short-dash).  For reference, a circle
inclined at $45^\circ$ is also shown by the dotted lines for spin.  Axes
are labeled in units of $M$ (corresponding to an angular scale of roughly
$5\,\muas$ for \SgrA).}
\label{cents_ISCO_350}
\end{figure}
The centroid paths for orbits at the ISCO of the three black hole
spins we considered are shown in Figure \ref{cents_ISCO_350}.  As in
the NIR, the inner orbits are substantially modified by strong
lensing, and in this case, opacity.  However, as mentioned previously,
the flux variation associated with the high-spin case is small, and
thus it is not clear that the centroid path will be measurable in this
case.

\section{Conclusions} \label{C}

The photon trapping radius of the black hole in the center of the
Milky-Way galaxy has the largest apparent sizes among all known black
holes, occupying an anglular scale of tens of micro-arcseconds on the
sky. This scale will be within reach of forthcoming observatories, such as
a sub-millimetre VLBA (for imaging), or the PRIMA instrument on the
VLT (for monitoring shifts in the infrared image centroid during flares).
Imaging hot spots in the immediate vicinity of an accreting black hole
provides a new method for testing general relativity and measuring the
black hole mass and spin.

The image and centroid shifts of a compact population of non-thermal
electrons may be used to constrain the mass and spin of the central
black hole.
At
radio wavelengths such a spot would persist for many orbitally
periods, ultimately being sheared into a ring, while in the NIR such a
spot would cool rapidly, surviving approximately a single orbit.

At the sub-millimetre frequencies currently proposed for the VLBA
imaging of the Galactic centre, the opacity of the underlying steady
accretion flow has a significant effect upon the images and centroid
motions.  Nonetheless, it remains possible to extract the black hole
mass and spin from these.  Furthermore, combined with the NIR data (at
which the accretion flow opacity is negligble) this will provide a
method by which the accretion flow may be isolated.

While polarization is significant in both the radio and NIR regimes,
the enhanced opacity at radio wavelengths reduces the polarization
signatures of strong lensing that are most noticeable for the
polarization angle and polarized flux in the NIR.  Immediately prior
to maximum magnification, rapid changes in the orientation of the
polarization take place in the NIR.  These are systematically smoothed
in the radio, though the polarization does rotate as expected for a
toroidal magnetic field.


We have used simple models for the underlying accretion flow and the
hot spot.  Future work can improve upon our results using
three-dimensional, fully relativistic, magneto-hydrodynamic
simulations of the accretion flow, in which hot spots (or flares)
arise naturally as magnetic reconnection events and/or at shocks.

\section*{Acknowledgments}
This work was supported in part by NASA grant NAG 5-13292 and NNG05GH54G
(for A.L.).  A.E.B. gratefully acknowledges the support of an ITC
Fellowship from Harvard College Observatory.

\bibliographystyle{mn2e.bst}
\bibliography{spot.bib}

\begin{thebibliography}{}

\bibitem[\protect\citeauthoryear{{Baganoff}, {Bautz}, {Brandt}, {Chartas},
  {Feigelson}, {Garmire}, {Maeda}, {Morris}, {Ricker}, {Townsley} \&
  {Walter}}{{Baganoff} et~al.}{2001}]{Baga_etal:01}
{Baganoff} F.~K.,  {Bautz} M.~W.,  {Brandt} W.~N.,  {Chartas} G.,  {Feigelson}
  E.~D.,  {Garmire} G.~P.,  {Maeda} Y.,  {Morris} M.,  {Ricker} G.~R.,
  {Townsley} L.~K.,    {Walter} F.,  2001, \nat, 413, 45

\bibitem[\protect\citeauthoryear{{Broderick} \& {Blandford}}{{Broderick} \&
  {Blandford}}{2003}]{Brod-Blan:03}
{Broderick} A.,  {Blandford} R.,  2003, \mnras, 342, 1280

\bibitem[\protect\citeauthoryear{{Broderick} \& {Blandford}}{{Broderick} \&
  {Blandford}}{2004}]{Brod-Blan:04}
{Broderick} A.,  {Blandford} R.,  2004, \mnras, 349, 994

\bibitem[\protect\citeauthoryear{{Broderick}}{{Broderick}}{2005}]{Brod:05}
{Broderick} A.~E.,  2005, \mnras, submitted

\bibitem[\protect\citeauthoryear{{Broderick} \& {Loeb}}{{Broderick} \&
  {Loeb}}{2005a}]{Brod-Loeb:05b}
{Broderick} A.~E.,  {Loeb} A.,  2005a, \apjl, submitted (astro-ph/0508386)

\bibitem[\protect\citeauthoryear{{Broderick} \& {Loeb}}{{Broderick} \&
  {Loeb}}{2005b}]{Brod-Loeb:05}
{Broderick} A.~E.,  {Loeb} A.,  2005b, \mnras, in press (astro-ph/0506433)

\bibitem[\protect\citeauthoryear{{Connors}, {Stark} \& {Piran}}{{Connors}
  et~al.}{1980}]{Conn-Star:80}
{Connors} P.~A.,  {Stark} R.~F.,    {Piran} T.,  1980, \apj, 235, 224

\bibitem[\protect\citeauthoryear{{De Villiers}, {Hawley} \& {Krolik}}{{De
  Villiers} et~al.}{2003}]{DeVi-Hawl-Krol:03}
{De Villiers} J.,  {Hawley} J.~F.,    {Krolik} J.~H.,  2003, \apj, 599, 1238

\bibitem[\protect\citeauthoryear{{Doeleman} \& {Bower}}{{Doeleman} \&
  {Bower}}{2004}]{Doel-Bowe:04}
{Doeleman} S.,  {Bower} G.,  2004, Galactic Center Newsletter, 18, 6

\bibitem[\protect\citeauthoryear{{Eckart}, {Baganoff}, {Morris}, {Bautz},
  {Brandt}, {Garmire}, {Genzel}, {Ott}, {Ricker}, {Straubmeier}, {Viehmann},
  {Sch{\" o}del}, {Bower} \& {Goldston}}{{Eckart} et~al.}{2004}]{Ecka_etal:04}
{Eckart} A.,  {Baganoff} F.~K.,  {Morris} M.,  {Bautz} M.~W.,  {Brandt} W.~N.,
  {Garmire} G.~P.,  {Genzel} R.,  {Ott} T.,  {Ricker} G.~R.,  {Straubmeier} C.,
   {Viehmann} T.,  {Sch{\" o}del} R.,  {Bower} G.~C.,    {Goldston} J.~E.,
  2004, \aap, 427, 1


\bibitem[\protect\citeauthoryear{{Falcke}, {Melia} \& {Agol}}{{Falcke}
  et~al.}{2000}]{Falc-Meli-Agol:00}
{Falcke} H.,  {Melia} F.,    {Agol} E.,  2000, \apjl, 528, L13

\bibitem[\protect\citeauthoryear{{Genzel}, {Sch{\" o}del}, {Ott}, {Eckart},
  {Alexander}, {Lacombe}, {Rouan} \& {Aschenbach}}{{Genzel}
  et~al.}{2003}]{Genz_etal:03}
{Genzel} R.,  {Sch{\" o}del} R.,  {Ott} T.,  {Eckart} A.,  {Alexander} T.,
  {Lacombe} F.,  {Rouan} D.,    {Aschenbach} B.,  2003, \nat, 425, 934

\bibitem[\protect\citeauthoryear{{Ghez}, {Wright}, {Matthews}, {Thompson}, {Le
  Mignant}, {Tanner}, {Hornstein}, {Morris}, {Becklin} \& {Soifer}}{{Ghez}
  et~al.}{2004}]{Ghez_etal:04}
{Ghez} A.~M.,  {Wright} S.~A.,  {Matthews} K.,  {Thompson} D.,  {Le Mignant}
  D.,  {Tanner} A.,  {Hornstein} S.~D.,  {Morris} M.,  {Becklin} E.~E.,
  {Soifer} B.~T.,  2004, \apjl, 601, L159

\bibitem[\protect\citeauthoryear{{Goldwurm}, {Brion}, {Goldoni}, {Ferrando},
  {Daigne}, {Decourchelle}, {Warwick} \& {Predehl}}{{Goldwurm}
  et~al.}{2003}]{Gold_etal:03}
{Goldwurm} A.,  {Brion} E.,  {Goldoni} P.,  {Ferrando} P.,  {Daigne} F.,
  {Decourchelle} A.,  {Warwick} R.~S.,    {Predehl} P.,  2003, \apj, 584, 751

\bibitem[\protect\citeauthoryear{{Jones} \& {O'Dell}}{{Jones} \&
  {O'Dell}}{1977}]{Jone-ODel:77a}
{Jones} T.~W.,  {O'Dell} S.~L.,  1977, \apj, 214, 522

\bibitem[\protect\citeauthoryear{{Laor}, {Netzer} \& {Piran}}{{Laor}
  et~al.}{1990}]{Laor-Netz-Pira:90}
{Laor} A.,  {Netzer} H.,    {Piran} T.,  1990, \mnras, 242, 560

\bibitem[\protect\citeauthoryear{{Lindquist}}{{Lindquist}}{1966}]{Lind:66}
{Lindquist} R.~W.,  1966, Anals of Physics, 37, 487

\bibitem[\protect\citeauthoryear{{Miyoshi}, {Ishitsuka}, {Kameno}, {Shen} \&
  {Horiuchi}}{{Miyoshi} et~al.}{2004}]{Miyo_etal:04}
{Miyoshi} M.,  {Ishitsuka} J.~K.,  {Kameno} S.,  {Shen} Z.,    {Horiuchi} S.,
  2004, Progress of Theoretical Physics Supplement, 155, 186

\bibitem[\protect\citeauthoryear{{Paumard}, {Perrin}, {Eckart}, {Genzel},
  {L\'ena}, {Sch\"odel}, {Eisenhauer}, {M\"uller} \& {Gillessen}}{{Paumard}
  et~al.}{2005}]{Paum_etal:05}
{Paumard} T.,  {Perrin} G.,  {Eckart} A.,  {Genzel} R.,  {L\'ena} P.,
  {Sch\"odel} R.,  {Eisenhauer} F.,  {M\"uller} T.,    {Gillessen} S.,  2005
  "ESO Astrophysics Symposia", "{S}cientific prospects for vlti in the galactic
  centre: Getting to the schwarzschild radius" {\em in press}

\bibitem[\protect\citeauthoryear{{Takahashi}}{{Takahashi}}{2004}]{Taka:04}
{Takahashi} R.,  2004, \apj, 611, 996

\bibitem[\protect\citeauthoryear{{Takahashi}}{{Takahashi}}{2005}]{Taka:05}
{Takahashi} R.,  2005, \pasj, 57, 273

\bibitem[\protect\citeauthoryear{{Yuan}, {Quataert} \& {Narayan}}{{Yuan}
  et~al.}{2003}]{Yuan-Quat-Nara:03}
{Yuan} F.,  {Quataert} E.,    {Narayan} R.,  2003, \apj, 598, 301

\bibitem[\protect\citeauthoryear{{Yuan}, {Quataert} \& {Narayan}}{{Yuan}
  et~al.}{2004}]{Yuan-Quat-Nara:04}
{Yuan} F.,  {Quataert} E.,    {Narayan} R.,  2004, \apj, 606, 894

\end{thebibliography}

\bsp

\end{document}